
\def\aap #1 #2 #3 {{\sl Adv.\ Appl.\ Prob.} {\bf #1}, #2 (#3)}
\def\acp #1 #2 #3 {{\sl Adv.\ Chem.\ Phys.} {\bf #1}, #2 (#3)}
\def\japr #1 #2 #3 {{\sl J.\ Appl.\ Prob.} {\bf #1}, #2 (#3)}
\def\jcp #1 #2 #3 {{\sl J.\ Chem.\ Phys.} {\bf #1}, #2 (#3)}
\def\jpa #1 #2 #3 {{\sl J. Phys.\ A} {\bf #1}, #2 (#3)}
\def\jsp #1 #2 #3 {{\sl J. Stat.\ Phys.} {\bf #1}, #2 (#3)}
\def\zw #1 #2 #3 {{\sl Z. Wahrsch.\ verw.\ Gebiete} {\bf #1}, #2 (#3)}

\def\gl{\mathrel{\raise1ex\hbox{$>$\kern-.75em\lower1ex\hbox{$<$}}}}
\def\lg{\mathrel{\raise1ex\hbox{$<$\kern-.75em\lower1ex\hbox{$>$}}}}
\def\gtwid{\mathrel{\raise.3ex\hbox{$>$\kern-.75em\lower1ex\hbox{$\sim$}}}}
\def\ltwid{\mathrel{\raise.3ex\hbox{$<$\kern-.75em\lower1ex\hbox{$\sim$}}}}
\def\sqr#1#2{{\vcenter{\hrule height.#2pt
      \hbox{\vrule width.#2pt height#1pt \kern#1pt
         \vrule width.#2pt}
      \hrule height.#2pt}}}

\def\cxt{\hbox{{$c(x,t)$}}}
\def\intt{\int_0^t}
\overfullrule=0pt

\def\eg{\hbox{{\it e.\ g.}}}\def\ie{\hbox{{\it i.\ e.}}}


\def\leaderfill{\leaders\hbox to 1em{\hss.\hss}\hfill}

\def\CA{\hbox{{$\cal A$}}}
\def\CB{\hbox{{$\cal B$}}}
\def\CC{\hbox{{$\cal C$}}}
\def\CD{\hbox{{$\cal D$}}}

\def\CF{\hbox{{$\cal F$}}}

\def\ref#1{${}^{#1}$}
\newcount\eqnum \eqnum=0  
\newcount\eqnA\eqnA=0\newcount\eqnB\eqnB=0\newcount\eqnC\eqnC=0\newcount\eqnD\eqnD=0
\def\eqnoi{\global\advance\eqnum by 1\eqno(\the\eqnum)}
\def\eqnai{\global\advance\eqnum by 1\eqno(\the\eqnum{a})}
\def\eqnbi{\eqno(\the\eqnum{b})}
\def\eqnoA{\global\advance\eqnA by 1\eqno(A\the\eqnA)}
\def\eqnoB{\global\advance\eqnB by 1\eqno(B\the\eqnB)}
\def\eqnoC{\global\advance\eqnC by 1\eqno(C\the\eqnC)}
\def\eqnoD{\global\advance\eqnD by 1\eqno(D\the\eqnD)}
\def\back#1{{\advance\eqnum by-#1 Eq.~(\the\eqnum)}}
\def\backs#1{{\advance\eqnum by-#1 Eqs.~(\the\eqnum)}}
\def\backn#1{{\advance\eqnum by-#1 (\the\eqnum)}}
\def\backA#1{{\advance\eqnA by-#1 Eq.~(A\the\eqnA)}}
\def\backB#1{{\advance\eqnB by-#1 Eq.~(B\the\eqnB)}}
\def\backC#1{{\advance\eqnC by-#1 Eq.~(C\the\eqnC)}}
\def\backD#1{{\advance\eqnD by-#1 Eq.~(D\the\eqnD)}}
\def\last{{Eq.~(\the\eqnum)}}                   
\def\lasts{{Eqs.~(\the\eqnum)}}                   
\def\lastn{{(\the\eqnum)}}                      
\def\lastA{{Eq.~(A\the\eqnA)}}\def\lastB{{Eq.~(B\the\eqnB)}}
\def\lastC{{Eq.~(C\the\eqnC)}}\def\lastD{{Eq.~(D\the\eqnD)}}
\newcount\refnum\refnum=0  
\def\refi{\smallskip\global\advance\refnum by 1\item{\the\refnum.}}

\newcount\rfignum\rfignum=0  
\def\rfigi{\medskip\global\advance\rfignum by 1\item{Figure \the\rfignum.}}

\newcount\fignum\fignum=0  
\def\figi{\global\advance\fignum by 1 Fig.~\the\fignum}

\newcount\rtabnum\rtabnum=0  
\def\rtabi{\medskip\global\advance\rtabnum by 1\item{Table \the\rtabnum.}}

\newcount\tabnum\tabnum=0  
\def\tabi{\global\advance\tabnum by 1 Table~\the\tabnum}

\newcount\secnum\secnum=0 
\def\chap#1{\global\advance\secnum by 1
\bigskip\centerline{\bf{\the\secnum}. #1}\smallskip\noindent}

\def\pd#1#2{{\partial #1\over\partial #2}}      
\def\p2d#1#2{{\partial^2 #1\over\partial #2^2}} 
\def\pnd#1#2#3{{\partial^{#3} #1\over\partial #2^{#3}}} 
\def\td#1#2{{d #1\over d #2}}      
\def\t2d#1#2{{d^2 #1\over d #2^2}} 
\def\av#1{\langle #1\rangle}                    


\def\2kth{{$2k^{\rm th}$}}

\def\n-th{{$(n-1)^{\rm th}$}}

\def\N-th{{$(N-1)^{\rm th}$}}

\def\0th{$0^{\rm th}$}
\def\1st{$1^{\rm st}$}
\def\2nd{$2^{\rm nd}$}
\def\3rd{$3^{\rm rd}$}
\def\4th{$4^{\rm th}$}
\def\5th{$5^{\rm th}$}
\def\5th{$6^{\rm th}$}
\def\6th{$7^{\rm th}$}
\def\7th{$7^{\rm th}$}
\def\8th{$8^{\rm th}$}
\def\9th{$9^{\rm th}$}

\def\a{{\alpha}}
\def\b{{\beta}}

\def\d{\delta}

\def\n{\nu}

\def\p{\pi}

\def\t{\tau}

\magnification=1200

\centerline{\bf Life and Death in a Cage and at the Edge of a Cliff}
\vskip 0.25in
\centerline{P.~L.~Krapivsky and S.~Redner}
\bigskip
\centerline{\sl Center for Polymer Studies and Department of Physics}
\centerline{\sl Boston University, Boston, MA 02215}
\vskip 0.4in

\centerline{\bf Abstract}\bigskip\bigskip

{
\narrower\narrower

\noindent  The survival probabilities of a ``prisoner'' diffusing in an
expanding cage and a ``daredevil'' diffusing at the edge of a receding
cliff are investigated.  When the diffuser reaches the boundary, he dies.
For ``marginal'' boundary motion, \ie, the cage length grows as
$\sqrt{At}$ or the cliff location recedes as $x_0(t)=-\sqrt{At}$ and the
daredevil diffuses within the domain $x>x_0$, the survival probability
of the diffuser exhibits non-universal power-law behavior, $S(t)\sim
t^{-\b}$, which depends on the relative rates of boundary and diffuser
motion.  Heuristic approaches are applied for the cases of ``slow'' and
``fast'' boundary motion which yield approximate expressions for $\b$.
An asymptotically exact analysis of these two problems is also performed
and the approximate expressions for $\b$ coincide with the exact results
for nearly entire range of possible boundary motions.

}

\vskip 0.4in
\baselineskip=20 truebp
\chap{Introduction}

Consider an inebriated ``prisoner'' who diffuses within a
one-dimensional cage defined by $(-L(t),L(t))$ and dies whenever he
touches the walls (Fig.~1).  We are interested in determining the
probability for such a prisoner to survive until time $t$, $S(t)$.
Mathematically, the time evolution of this process in governed by the
diffusion equation
$$
\pd{c(x,t)} t = D\pnd{c(x,t)} x 2,
\eqnoi
$$
within the domain $-L(t)\le x \le L(t)$, subject to the initial
condition $c(x,t=0)=\d(x)$, and the absorbing boundary condition
$c(x=\pm L(t),t)=0$.  Here $\cxt$ is the prisoner density at position
$x$ and time $t$, and $D$ is the diffusion coefficient.  The absorbing
boundary condition imposes the death of the prisoner whenever he touches
the walls of the cage.  Correspondingly, the survival probability $S(t)$
is given by the spatial integral of the prisoner density,
$$
S(t)\equiv \int_{-L(t)}^{L(t)}\cxt\,dx. \eqnoi
$$

In cage of fixed size $2L$, the survival probability
decays as $\exp(-\pi^2Dt/4L^2)$ in the long-time limit [1].  More
interesting behavior arises when we aid the prisoner by allowing the
cage walls to recede.  This obviously increases the prisoner lifetime
and can also dramatically change the form of the prisoner survival
probability.  In the following, we consider power-law growth of the cage
length, $L(t)\simeq (At)^\a$, as $t\to\infty$.  Two domains of behavior
arise which are determined by the relative magnitudes of the basic
length scales of the system, namely, the cage length $2L(t)$ and the
diffusion length $\sqrt{Dt}$.  When $L(t) \ll \sqrt{Dt}$, which arises
when $\a<1/2$ or when $\a=1/2$ and $A\ll D$, the prisoner diffuses
faster than the cage walls recede and it is plausible to apply an
approximation based on the assumption that the prisoner probability
density is ``close'' to that in the case where the cage walls are
static.  This is the basis of the well-known adiabatic approximation
[2].  Conversely, for a rapidly expanding cage, \ie, $\a>1/2$ or
$\a=1/2$ and $A\gg D$, the cage expands faster than the prisoner
diffuses and approximations based on a ``free'' prisoner should be
appropriate.

In the interesting situation where the diffusion length of the prisoner
and the cage size are comparable, both the adiabatic and the ``free
prisoner'' approximations predict that the prisoner survival probability
decays as a power law in time, $S(t)\sim t^{-\b}$, but with a {\it
non-universal\/} exponent which depends on $A/D$.  To ascertain the
accuracy of these heuristic approaches, we analyze of the diffusion
equation with the moving boundary condition and find that the prisoner
probability density may be written in terms of parabolic cylinder
functions.  From this analysis, we determine the exponent $\b$ for all
$A/D$ and verify that the results of the above two heuristic approaches
are asymptotically exact.

We also consider the related problem of an inebriated ``daredevil'' who
diffuses in a one-dimensional semi-infinite domain $x>x_0(t)$, and fall
to his death whenever $x=x_0(t)$ is reached.  When the cliff location is
fixed, it is well-known that the survival probability of the daredevil
is $S(t)\equiv\int_{x_0}^\infty \cxt\,dt\sim t^{-1/2}$ [1].  Thus the
daredevil is sure to fall off the cliff (although his mean lifetime is
infinite).  This same decay law continues to hold if the cliff recedes
slowly, \ie, $x_0\sim (At)^\a$ with $\a<1/2$ [3].  Conversely, if the
cliff recedes from the daredevil at a constant velocity $v$, there is
finite probability for the daredevil to survive which rapidly approaches
unity when $v\ell_0/D>1$.  Here $\ell_0$ is the initial distance from
the daredevil to the cliff.  However, if the cliff recedes at the same
rate as the daredevil diffuses, $x_0(t)\sim -\sqrt{At}$ with $A$ of the
order of $D$, marginal behavior again arises in which the daredevil
survival probability exhibits a non-universal power-law decay in time.

It is worth mentioning that the first-passage probability of diffusion
processes in the presence of moving, absorbing boundaries has been
investigated previously by mathematicians and a substantial literature
exists (see \eg, [3-5] and references therein).  However, the methodology
used in these papers is quite different from the simple-minded approach
that we will present.  In addition to differences in approach, an
advantageous aspect of our study is that the prisoner and daredevil
problems can be treated within the same framework.

\chap{Heuristic Approaches for Prisoner Survival in an Expanding Cage}

For a fixed  cage $(-L,L)$, the solution to the  diffusion
equation (1) may be written as an eigenfunction expansion in which each
eigenmode decays exponentially in time, with a different characteristic
decay time.  In the long time limit, only the most slowly decaying
eigenmode remains and the the density approaches
$$
\cxt \propto e^{-Dt/4L^2}\,\,\cos\left(\pi x \over 2L\right).
\eqnoi
$$
Thus the prisoner survival probability decays exponentially in time.

Now suppose that the cage expands expands slowly, $L(t)\ll\sqrt{Dt}$.
In this case, the adiabatic approximation shows that the density profile
approaches the same form as in the fixed-cage case, except that the
parameters in this probability distribution acquire time dependence to
satisfy the moving boundary condition.  The corresponding
probability density is
$$
\cxt \propto f(t)\,\cos\left(\pi x \over 2L(t)\right).
\eqnoi
$$
with the amplitude $f(t)$ to be determined. Substituting \last\ into
Eq.~(1) leads to
$$
\dot f = -\left(D\pi^2 \over 4L^2\right) f - \left(\pi x\over
2L^2\right) \tan({\pi x\over 2L})\,\, \dot L f.
\eqnoi
$$
When $L(t)$ grows as $(At)^\a$ with $\a<1/2$, the second term on the
right-hand side may be neglected and the
leading behavior of the amplitude $f(t)$ is given by
$$
f(t) \to \exp\left[-{\pi^2 D\over 4} \intt\,dt'\,
L^{-2}(t')\right].
\eqnoi
$$
The full asymptotic behavior of $f(t)$ is expected to also contain a
power-law prefactor in $L$.  However, this prefactor is not accessible
within our naive approach.  The leading behavior of $f(t)$ now gives
$$
S(t)=\int_{-L(t)}^{L(t)} ~\cxt\,dx \simeq ~{4\over \pi}~f(t)L(t).
\eqnoi
$$
Thus for $\a<1/2$ the leading behavior of the survival probability decays as a
stretched exponential in time
$$
S(t)  \to\exp\left[-{\pi^2 D\over 4(1-2\a)A^{2\a}}t^{1-2\a}\right].
\eqnoi
$$
For the marginal case of $\a=1/2$, the second term in \back3\ is no
longer negligible.  Following the above procedure nevertheless,
we find a non-universal power-law behavior,
$$
S(t) \sim t^{-\b},\quad {\rm with}
\quad \b={\pi^2 D\over 4A}.
\eqnoi
$$
Given the nature of the approximation employed, this prediction is
anticipated to be valid in the limit  $A\ll D$.

In the complementary case of a rapidly growing cage, $L(t) \gg
\sqrt{Dt}$, \ie, $\a>1/2$ or $\a=1/2$ and $A\gg D$, it is plausible
to assume that the prisoner density profile approaches a Gaussian but
with a decaying overall integral.  Thus we are led to hypothesize
$$
\cxt \simeq {S(t)\over \sqrt{4\pi Dt}}\exp\left(-{x^2\over 4Dt}\right).
\eqnoi
$$
Although this distribution does not satisfy the absorbing boundary
condition, the inconsistency is expected to be negligible, since the
density is exponentially small at the cage walls.  The decay of
the mass may be found by equating the flux to the cage walls,
$2j=-2D\pd c x$, with the mass loss.  An
elementary computation shows that the survival probability
approaches a constant for $\a>1/2$.
In the marginal case of $\a=1/2$,
$$
\dot S  = -{S\over t}\sqrt{A \over 4\pi D}\exp\left(-{A\over 4D}\right).
\eqnoi
$$
This again leads to power law behavior for the survival
probability, $S\sim t^{-\b}$
with
$$
\b=\sqrt{A \over 4\pi D}\exp\left(-{A\over 4D}\right).
\eqnoi
$$
The predictions of this ``free prisoner'' approximation should be
accurate when $A\gg D$.  As is intuitively expected, when $A/D$
decreases toward 0, $\b$ diverges.  This corresponds to the survival
probability exhibiting faster than power-law decay, as predicted by the
adiabatic approximation.  Conversely, as $A/D\to\infty$ the prisoner
diffuses more slowing than the cage expands and there is a finite chance
to survive asymptotically, \ie, $\b\to 0$.

\chap{Asymptotic Analysis for the Marginally Growing Cage}

Let us now investigate more carefully the borderline case where the
cage grows as $L(t)\simeq\sqrt{At}$.  Within a scaling
formulation, it is natural to hypothesize that
the density can be written in terms of the dimensionless variables
$$
\xi\equiv{x\over L(t)},\qquad \sigma\equiv {x\over\sqrt{Dt}},
$$
as $t\to\infty$.
Since both the cage length and the diffusion length grow at the same
rate, it proves convenient to consider the basic variables to be $\xi$ and
$\rho=\xi/\sigma= A/D$ and write the concentration as
$$
\cxt \sim t^{-\b-1/2}\CC_\rho(\xi).
\eqnoi
$$
The power law prefactor is chosen to ensure that the survival
probability decays as $t^{-\b}$, as defined previously.

Substituting \last\ into Eq.~(1), we find that the scaling function
$\CC_{\rho}(\xi)$ satisfies the ordinary differential equation
$$
{D\over A}{d^2\CC \over d\xi^2}
+{1\over 2}\xi\td\CC \xi+
\left(\b+{1\over 2}\right)\CC=0,
\eqnoi
$$
where the $\rho$ dependence has been dropped, for notational simplicity.
Introducing $\xi=\eta\,\sqrt{2D/A}$ and
$\CC(\xi)=e^{-\eta^2/4}\,\CD(\eta)$, transforms \last\ into the
canonical form for the parabolic cylinder equation [6]
$$
{d^2\CD \over
d\eta^2} + \left(2\b+{1\over 2}-{\eta^2\over 4}\right)\CD=0.
\eqnoi
$$
A spatially symmetric solution to \last\ (appropriate for the prisoner
starting in the middle of the cage) is
$$
\CD(\eta)={1\over 2}(\CD_{2\b}(\eta)+\CD_{2\b}(-\eta)), \eqnoi
$$
with $\CD_\nu(\eta)$ the parabolic cylinder function of order $\nu$.
Finally, the relation between the decay exponent $\b$ and $A/D$ is
determined by the absorbing boundary condition
$$
\CD_{2\b}\left(\sqrt{A\over 2D}\right)+
\CD_{2\b}\left(-\sqrt{A\over 2D}\right)=0.
\eqnoi
$$

One can easily determine the limiting behaviors of $\b=\b(A/D)$ for $A/D\to
0$ and $A/D\to \infty$ and thus check the validity of the heuristic
predictions given in \backs{8} and \backn5, respectively.  In
the former case, the exponent $\b$ is large and hence, from \back2, the
profile approaches the cosine form.  Thus \back{8} provides the correct
asymptotics.  In the
latter case, $A\gg D$, $\b\to 0$ and \back2 approaches the Schr\"odinger
equation for the ground state of the harmonic oscillator, for which
$\CD_0(\eta)=\exp(-\eta^2/4)$.  For small but finite $\b$ it is natural
to seek a perturbative solution
$$
\CD(\eta)=\exp(-\eta^2/4)+\b\CA(\eta)+\ldots.
\eqnoi
$$
Substituting this expansion into \back3 yields an inhomogeneous
linear equation for the correction $\CA(\eta)$:
$$
{d^2\CA \over d\eta^2} + \left({1\over 2}-{\eta^2\over 4}\right)\CA
=-2e^{-\eta^2/4}.
\eqnoi
$$
Introducing $\CB(\eta)$ through $\CA(\eta)=\exp(-\eta^2/4)\CB(\eta)$, we obtain
$\CB''-\eta\CB'=-2$.  By solving this latter equation,
the perturbative solution for $\CD(\eta)$ is
$$
\CD(\eta)=e^{-\eta^2/4}-2\b e^{-\eta^2/4}
\int_0^{\eta}d\eta_1\,e^{\eta_1^2/2}\int_0^{\eta_1}\,d\eta_2\,e^{-\eta_2^2/2}+O(\b^2).
\eqnoi
$$
Combining \last\ and the absorbing boundary condition
$\CD(\sqrt{A/2D})=0$ we reproduce (after a straightforward but lengthy
computation) the asymptotics given by \back{8}.  Thus we have
rigorously justified the previous heuristic predictions
predictions for the decay exponent $\b$.

It is also interesting to consider the mean lifetime of the prisoner,
$\av{t}=\int_0^\infty dt\, S(t)$.  This quantity is finite for $\b>1$
and infinite for $\b<1$.  The borderline case of $\b=1$, corresponds to
the second excited state of the wavefunction in the harmonic oscillator
potential in \back5, for which the solution for $\CD(\eta)$ is
$\CD(\eta)=(1-\eta^2)e^{-\eta^2/4}$.  The boundary condition of \back3
now gives $A=2D$.  Thus the borderline case between a finite and an
infinite mean survival time corresponds to $A=2D$.  It is gratifying
that for this case of $A=2D$, the simple-minded adiabatic approach gives
$\b \cong 1.234$.  This provides a sense for the accuracy of the
adiabatic approach in the regime $A<2D$.

\chap{Survival of a Daredevil at the Edge of a Cliff}

Let us now turn to the case of an inebriated daredevil who survives if
he remains within the semi-infinite domain $x_0(t)\le x <\infty$.  As
mentioned in the introduction, the most interesting case is where
$x_0(t)\simeq-\sqrt{At}$ as $t\to \infty$, corresponding to the cliff
receding at the same rate at which diffusion tends to transport the
daredevil to the cliff.  This situation can be analyzed by methods
similar to those applied for the prisoner in the marginally growing
cage.

It is convenient to first change variables from $(x,t)$ to
$(x'=x-x_0(t),t)$ to fix the absorbing boundary at the origin.  Thus the
initial diffusion equation is transformed to the convection-diffusion
equation (where the prime is now dropped)
$$
\pd c t - {x_0\over 2t}\pd c x= D\pnd c x 2,
\quad {\rm for} \quad 0\le x <\infty,
\eqnoi
$$
with the {\it fixed\/} absorbing boundary condition $c(x=0,t)=0.$   As in
the case of the prisoner in the marginally expanding cage, we apply the
same scaling assumption, \back{8}, for the probability
density of the daredevil.  Substituting this form into \last\ gives
$$
{D\over A}{d^2\CC \over d\xi^2} +{1\over 2}(\xi-1)\td\CC \xi+
\left(\b+{1\over 2}\right)\CC=0.
\eqnoi
$$
Transforming from $\xi=-{x\over x_0}$ and $\CC(\xi)$ to $\eta$ and
$\CD(\eta)$ in a slightly different way than previously,
$$
\xi-1=\sqrt{2D\over A}~\eta,\quad \CC(\xi)=\exp\left(-{\eta^2\over
4}\right)\CD(\eta),
\eqnoi
$$
we find that $\CD(\eta)$ satisfies the same parabolic cylinder equation
(Eq.~(16)) as in the prisoner problem.  However, slightly
different boundary conditions apply.  Death of the daredevil at
the edge of the cliff implies
$$
\CD(-\sqrt{A/2D})=0.
\eqnai
$$
On the other hand,
$S(t)\equiv\int_0^\infty dx~\cxt \le 1$ implies the boundary
condition at $\eta=\infty$,
$$
\CD(\eta=\infty)=0.
\eqnbi
$$

Mathematically, the determination of $\b$ and $\CD(\eta)$ is equivalent
to finding the ground state energy and wavefunction of a quantum
particle in a potential composed of an infinite barrier at
$\eta=-\sqrt{A/2D}$ and the harmonic oscillator potential for
$\eta>-\sqrt{A/2D}$.  Higher excited  states do
not contribute in the long time limit.
A general solution to \back2 satisfying $\CD(\infty)=0$ is
$$
\CD(\eta)=\CD_{2\b}(\eta),
\eqnoi
$$
and the absorbing boundary condition $\CD_{2\b}(-\sqrt{A/2D})=0$
determines the relation between the decay exponent $\b$ and $A/D$.

Since \last\ provides only an implicit relation $\b=\b(A/D)$, it is
useful to determine the limiting behaviors of $\b$ for small and large
values of $A/D$. These relations can be derived directly from \back4 and
elementary facts about the quantum mechanics of the harmonic
oscillator, rather than relying on mathematical properties of the
parabolic cylinder functions.  In the limit of slowly moving boundary,
$A\ll D$, the wall is close to the origin in the $\eta$-variable.  When
the wall is exactly at the origin, the ground state of the truncated
potential is obviously the first excited state for pure harmonic
oscillator, namely, $2\b=1$ and $\CD(\eta)=\eta\exp(-\eta^2/4)$.  For
$A\ll D$, we therefore expect that $0<1-2\b\ll 1$.  This again suggests
the perturbative solution,
$$
\CD(\eta)=\eta\exp(-\eta^2/4)+(1-2\b)\CA(\eta)+\ldots.
\eqnoi
$$
Substituting this expansion into the parabolic cylinder equation
yields for $\CA(\eta)$:
$$
{d^2\CA \over d\eta^2} + \left({3\over 2}-{\eta^2\over 4}\right)\CA
=\eta e^{-\eta^2/4}.
\eqnoi
$$
Introducing $\CB(\eta)$ through $\CA(\eta)=\eta\,\exp(-\eta^2/4)\,
\CB(\eta)$ we find $\CB''-(\eta-2/\eta)\CB'=1$.
Solving this equation subject to
the boundary condition Eq.~(24b) gives $\CB(\eta)$, from which one
ultimately obtains
$$
\CD(\eta)=\eta\, e^{-\eta^2/4}+(1-2\b)\,\eta\, e^{-\eta^2/4}
\int_{\eta}^{\infty} d\eta_1\, \eta_1^{-2}\,e^{\eta_1^2/2}
\int_{\eta_1}^{\infty}d\eta_2\, \eta_2^{2}\,e^{-\eta_2^2/2}+\ldots.
\eqnoi
$$
Applying the absorbing boundary condition Eq.~(24a) gives
$$
\b \simeq {1\over 2}-\sqrt{A \over 4\pi D}.
\eqnoi
$$

One can treat the opposite limit $A\gg D$ similarly.  In terms of the
coordinate $\eta$, the location of the wall goes to $-\infty$.  Hence the
unperturbed ground state for this system is just the ground state for
the pure harmonic oscillator, namely, $\b=0$ and $\CD(\eta)=\exp(-\eta^2/4)$.
Following the same perturbative approach as in the complementary case of
$A\ll D$, we find
$$
\CD(\eta)=e^{-\eta^2/4}+2\b\, e^{-\eta^2/4}
\int_{\eta}^{\infty}d\eta_1\,e^{\eta_1^2/2}
\int_{\eta_1}^{\infty}d\eta_2\,e^{-\eta_2^2/2}+O(\b^2).
\eqnoi
$$
Combining \last\ with the absorbing boundary condition one finds the
same expression for $\b$, Eq.~(12), as was found for a finite cage.

\chap{Summary and Discussion}

We have presented a heuristic and an asymptotically exact approach
to determine the survival probability and the density distribution for
(i) a prisoner in a growing cage, $(-L(t),L(t))$, and (ii) a daredevil
in the domain $x>x_0(t)$ with a cliff at $x=x_0(t)$.  We were primarily
concerned with the ``marginal'' case where $L(t)\cong\sqrt{At}$ and
$x_0\cong -\sqrt{At}$ (with $A$ of the order of $D$), so that boundary
of the system recedes at the same rate at which diffusion tends to bring
the diffuser (prisoner or daredevil) toward his demise.  In these
marginal situations, the survival probability of the diffuser exhibits a
non-universal power-law decay in time.  The value of the decay exponent
in the limiting cases of $A\gg D$ and $A\ll D$ can be obtained by simple
arguments.  These limiting behaviors are found to coincide with the
results from an asymptotic analysis of the underlying equation of
motion.

For the prisoner problem,  our results can be straightforwardly extended to
general spatial dimension.  Following the same adiabatic and
``free prisoner'' approximations that were applied in one dimension, we
find that the decay exponent $\b$ becomes
$$
\b=\cases{{j_d^2D\over A}  &\qquad adiabatic;  \cr
            {1\over\Gamma({d\over2})} \left(A\over{4D}\right)^{d/2}
\exp(-{A\over 4D}) & \qquad free. \cr}
\eqnoi
$$
Here $j_d$ is the first positive root of the spherical Bessel function
$J_{d/2-1}(x)$.
Similarly, in the case of marginal cage growth, the $d$-dimensional
analog of the scaling ansatz, Eq.~(13), can be applied, leading to a
generalization of Eq.~(14).
In three dimensions, in particular, the additional transformation
$\CC(\xi)=\CF(\xi)/\xi$ leads to the same parabolic cylinder equation
(14) for $\CF$, but with the parameter $\b+{1\over 2}$ replaced by
$\b+1$.

It is instructive to compare the behavior of the survival probabilities
given here with those of the related problem where the absorbing
boundaries themselves undergo diffusive motion.  These are situations
for which exact solutions have been given previously [7].  For example,
consider the survival of a diffusing daredevil when the position of the
cliff also diffuses with a diffusivity $A$.  This situation is
trivially isomorphic to the case of a static cliff and a daredevil with
diffusivity $D+A$. Thus the survival probability decays universally as
$t^{-1/2}$.  As might be expected, a cliff which is systematically
receding from a diffusing daredevil with $L(t)\propto\sqrt{t}$ leads to
a larger survival probability compared to the case of a stochastically
moving cliff.

On the other hand, the survival of a diffusing prisoner inside a cage
where both walls diffuse (each with diffusivity $A$) is more
interesting.  A variety of exact solutions show that the prisoner
survival probability decays non-universally, $S(t)\sim t^{-\b}$, with
decay exponent $\b=\pi/2\cos^{-1}(D/(D+A))$ [7].  For rapidly diffusing
walls, $\b\to 1$, while the corresponding limit for systematically
receding walls is $\b\to 0$.  Clearly if cage walls are receding
rapidly, the prisoner is more likely to survive compared to the case
where the cage walls are diffusing rapidly.  On the other hand, in the
limit of slowly diffusing walls, $\b\to \sqrt{\pi^2D\over{8A}}$.
Strangely, this is almost the square-root of the corresponding
expression for $\b$ quoted in Eq.~(9).  Intriguingly, the prisoner is
more likely to survive in a stochastically and slowly growing cage than
in a cage which grows systematically at the same average rate.

\chap{Acknowledgements}

We thank W. R. Young for an initial discussion on this problem and
G. H. Weiss for steering us towards the pertinent mathematical
literature.  We also gratefully acknowledge NSF grant DMR-9219845 for
partial support of this research.


\vfill\eject

\centerline{\bf References}\smallskip

\refi See \eg, G. H. Weiss and R. J. Rubin, \acp 52 363 1983 , and references
therein.

\refi See \eg, A. Messiah, {\sl Quantum Mechanics} (Interscience, New
York, 1961).

\refi K. Uchiyama, \zw 54 75 1980 .

\refi H. E. Daniels, \japr 6 399 1969 .

\refi P. Salminen, \aap 20 411 1988 .

\refi C.~M.~Bender and S.~A.~Orszag, {\sl Advanced Mathematical Methods
for Scientists and Engineers} (McGraw-Hill, New York, 1978);
M.~Abramowitz and I.~A.~Stegun, {\sl Handbook of  Mathematical
Functions}.  (Dover, New York, 1965).

\refi F. Leyvraz, (unpublished 1987); D. ben-Avraham, \jcp 88 941 1988 ;
M. E. Fisher and M. P. Gelfand, \jsp 53 175 1988 .


\vskip 0.4in

\centerline{\bf Figure Caption}\smallskip

\rfigi (a) The prisoner in the expanding cage, and (b) the daredevil at
the edge of a receding cliff.

\vfill\eject\bye